\documentclass[11pt,a4paper]{article}

\usepackage{footnote}

\usepackage[T1]{fontenc}
\usepackage[bitstream-charter]{mathdesign}

\usepackage[latin1]{inputenc}
\usepackage{amsmath}
\usepackage{subfig}
\usepackage{cite}

\usepackage{xcolor}
\definecolor{bl}{rgb}{0.0,0.2,0.6}

\usepackage{graphicx}
\usepackage{sectsty}
\usepackage[compact]{titlesec}
\allsectionsfont{\color{bl}\scshape\selectfont}

\makeatletter
\def\printtitle{%
    {\color{bl} \centering \huge \sc \textbf{\@title}\par}}
\makeatother

\title{\\ \large \vspace*{-10pt} Exact Spin and Pseudo-Spin Symmetric Solutions of the Dirac-Kratzer Problem with a tensor potential
via Laplace Transform Approach\vspace*{10pt}}

\makeatletter
\def\printauthor{%
    {\centering \small \@author}}
\makeatother

\author{%
    Altuð Arda \\
    arda@hacettepe.edu.tr \\
    Ramazan Sever \\
    sever@etu.edu.tr \\
    \vspace{20pt}
    }

\usepackage{fancyhdr}
\usepackage{lastpage}
\usepackage[runin]{abstract}
\abslabeldelim{\quad}
\catcode`ð=\active
 \defð{\u{g}}
 \catcode`Ð=\active
 \defÐ{\u{G}}
 \catcode`Ý=\active
 \defÝ{\. I}
 \catcode`ö=\active
 \defö{\"{o}}
 \catcode`Ö=\active
 \defÖ{\"O}
 \catcode`ü=\active
 \defü{\"{u}}
 \catcode`Ü=\active
 \defÜ{\"{U}}
 \catcode`Þ=\active
 \defÞ{\c{S}}
 \catcode`þ=\active
 \defþ{\c{s}}
 \catcode`ý=\active
 \defý{{\i}}
 \catcode`ç=\active
 \defç{\d{c}}
 \catcode`Ç=\active
 \defÇ{\d{C}}

\makesavenoteenv{tabular}
\begin{document}

\printtitle

\printauthor

\begin{abstract}
Exact bound state solutions of the Dirac equation for the Kratzer
potential in the presence of a tensor potential are studied by
using the Laplace transform approach for the cases of spin- and
pseudo-spin symmetry. The energy spectra is obtained in the closed
form for the relativistic as well as non-relativistic cases
including the Coulomb potential. It is seen that our analytical
results are in agrement with the ones given in literature. The
numerical results are also given in a table for different
parameter values.
\end{abstract}

\section{Introduction}
The analytical solvable problems within the framework of quantum
mechanics are very restricted. In this manner, it could be
interesting to find the exact analytical solutions of the Kratzer
potential in the relativistic domain. We intend to study the
radial Dirac equation for the potential having the form
\begin{eqnarray}
V(r)=-2Da\left(\frac{1}{r}-\frac{a/2}{r^2}\right)\,,
\end{eqnarray}
where $D>0$ and $a>0$. The detailed analysis of the bound states
of this potential including a third parameter $q$ in the
non-relativistic domain are placed in literature \cite{ey,rn,jac}
where used the WKB approximation and the semiclassical
quantization procedure with the help of the Langer transformation.
The potential form including the parameter $q$ provides the
extension of the spectra of the usual Kratzer potential
\cite{ak,ef}.

We find the exact bound state solutions and the corresponding
upper- and lower-spinor of the radial Dirac equation in the
presence of a tensor interaction within the framework of the
Laplace transformation (LT) scheme \cite{czd,gc1,gc2,aa} for the
cases of spin symmetry, $S(r)=+V(r)$, and pseudo-spin symmetry,
$S(r)=-V(r)$. The studying the solutions of the Dirac equation for
the spin and pseudo-spin symmetric cases has been received a great
attention in literature \cite{rlh,cyf,ad1}.

The organization of this work is as follows. In Section 2, we find
the energy spectrum and corresponding wave functions of the
Kratzer potential \cite{hh,ad2,rl,am} for the spin and pseudo-spin
symmetric cases which have been studied in details for the above
potential without the tensor interaction in Ref. \cite{or}. We
also study the energy spectra of the Coulomb potential and see
that the result obtained for the Coulomb problem is the same with
the one obtained in Ref. \cite{kth}. At the end, we give the bound
state spectrum of the above potentials for the non-relativistic
limit where the results are consistent with the ones obtained in
the literature.

\section{Bound States}

Dirac equation for the scalar $S(r)$ and vector $V(r)$ potentials
with a tensor interaction $U(r)$ is ($\hbar=c=1$) \cite{jng}
\begin{eqnarray}
\left\{\vec{\alpha}.\vec{p}+\beta\left[m_{0}+S(r)\right]+V(r)-i\beta.\hat{r}U(r)-E_{n\kappa}\right\}
\Psi(\vec{r})=0\,,
\end{eqnarray}
where $\vec{p}$ is the momentum operator, $E_{n\kappa}$ is the
relativistic energy of the particle with the rest mass $m_{0}$ and
$\vec{\alpha}$ written in terms of Pauli matrices and $\beta$ are
$4 \times 4$ Dirac matrices, respectively. The energy of the
particle $E_{n\kappa}$ depends on the radial quantum number $n$
and spin-orbit quantum number $\kappa$. Inserting the
eigenfunctions in terms of the spherical harmonic functions
$Y_{\kappa}(\theta,\phi)$ and $Y_{-\kappa}(\theta,\phi)$
\begin{eqnarray}
\Psi_{n \kappa}(r)=\,\frac{1}{r}\,\Bigg(\begin{array}{c}
  \,F(r)Y_{\kappa}(\theta,\phi) \\
iG(r)Y_{-\kappa}(\theta,\phi)
\end{array}\Bigg)\,,
\end{eqnarray}
gives the following coupled differential equations for the
upper-spinor $F(r)$ and lower-spinor $G(r)$
\begin{subequations}
\begin{align}
\left(\frac{d}{dr}+\frac{\kappa}{r}-U(r)\right)F(r)&=\left[m_{0}+E_{n\kappa}+S(r)-V(r)\right]G(r)\,,\\
\left(\frac{d}{dr}-\frac{\kappa}{r}+U(r)\right)G(r)&=\left[m_{0}-E_{n\kappa}+S(r)+V(r)\right]F(r)\,.
\end{align}
\end{subequations}
Substituting Eq. (4a) into Eq. (4b), we obtain a second-order
differential equation
\begin{eqnarray}
\left[\frac{d^2}{dr^2}-\frac{\kappa(\kappa+1)}{r^2}-[\frac{d}{dr}-\frac{2\kappa}{r}+U(r)]U(r)-
\frac{\frac{d}{dr}[S(r)-V(r)][\frac{d}{dr}+\frac{\kappa}{r}-U(r)]}{m_{0}+E_{n\kappa}+S(r)-V(r)}\right]F(r)\nonumber\\
-\left[m^2_{0}-E^2_{n\kappa}+2\left(m_{0}S(r)+E_{n\kappa}V(r)\right)+S^2(r)-V^2(r)\right]F(r)=0\,.
\end{eqnarray}

We consider the case where the scalar and vector potentials are
equal, $S(r)=V(r)$, which corresponds to the spin symmetry of the
Dirac equation \cite{rlh,cyf,ad1}. Inserting Eq. (1) into Eq. (5),
taking the tensor interaction as $-C/r$ with a constant $C$ and
using a new wave function as $F(r)=\sqrt{r\,}\phi(r)$ gives
\begin{eqnarray}
\frac{d^2\phi(r)}{dr^2}+\frac{1}{r}\frac{d\phi(r)}{r}-\frac{1}{r^2}\left(a^2_{1}+a^2_{2}r+
\epsilon^2_{n\kappa}r^2\right)\phi(r)=0\,,
\end{eqnarray}
where
\begin{subequations}
\begin{align}
a^2_{1}&=\kappa(\kappa+1)+C(1+2\kappa+C)+2Da^2(m_{0}+E_{n\kappa})+\frac{1}{4}\,,\\
a^2_{2}&=-4Da(m_{0}+E_{n\kappa})\,,\\
\epsilon^2_{n\kappa}&=-(E^2_{n\kappa}-m^2_{0})\,.
\end{align}
\end{subequations}
Defining a wave function $\phi(r)=r^{A}f(r)$ with $A$ as a
constant and inserting into Eq. (6) leads
\begin{eqnarray}
r\frac{d^2f(r)}{dr^2}-(2a_{1}-1)\frac{df(r)}{dr}-\left(a^2_{2}+\epsilon^2_{n\kappa}r\right)f(r)=0\,,
\end{eqnarray}
where we set $A=-a_{1}$ to obtain a finite solution when $r
\rightarrow \infty$. By using the LT defined as \cite{mrs}
\begin{eqnarray}
\mathcal{L}\left\{f(r)\right\}=f(t)=\int_{0}^{\infty}dr
e^{-tr}f(r)\,,
\end{eqnarray}
we obtain a first-order differential equation from Eq. (8) as
\begin{eqnarray}
\left(t^2-\epsilon^2_{n\kappa}\right)\frac{df(t)}{dt}+\left[\left(2a_{1}+1\right)t+a_{2}^2\right]f(t)=0\,.
\end{eqnarray}
It's solution is
\begin{eqnarray}
f(t)\sim \left(t+\epsilon_{n\kappa}\right)^{-(2a_{1}+1)}
\left(\frac{t-\epsilon_{n\kappa}}{t+\epsilon_{n\kappa}}\right)^{-\frac{a_{2}^2}
{2\epsilon_{n\kappa}}\,-\,\frac{2a_{1}+1}{2}}\,.
\end{eqnarray}
We impose the following condition to obtain a single-valued wave
function
\begin{eqnarray}
-\frac{a_{2}^2}
{2\epsilon_{n\kappa}}-\frac{1}{2}\,(2a_{1}+1)=n\,.\,\,\,(n=0, 1,
2, 3, \ldots)
\end{eqnarray}
Taking into account this condition and applying a simple series
expansion to Eq. (11) gives
\begin{eqnarray}
f(t)\sim
\sum_{m=0}^{n}\frac{(-1)^{m}n!}{(n-m)!m!}\,(2\epsilon_{n\kappa})^{m}\left(t+\epsilon_{n\kappa}\right)^{-(2a_{1}+1)-m}\,,
\end{eqnarray}
Using the inverse Laplace transformation \cite{mrs} in Eq. (13) we
obtain
\begin{eqnarray}
f(r) \sim r^{2a_{1}}e^{-\,\epsilon_{n\kappa}
r}\sum_{m=0}^{n}\frac{(-1)^{m}n!}{(n-m)!m!}\frac{\Gamma(2a_{1}+1)}{\Gamma(2a_{1}+1+m)}\,(2\epsilon_{n\kappa}
r)^{m}\,,
\end{eqnarray}
and
\begin{eqnarray}
F(r)=Nr^{a_{1}+\frac{1}{2}}e^{-\,\epsilon_{n\kappa}
r}\,_{1}F_{1}(-n,2a_{1}+1,2\epsilon_{n\kappa} r)\,,
\end{eqnarray}
where $N$ is a normalization constant and used the following
identity for the hypergeometric functions \cite{ma}
\begin{eqnarray}
_{1}F_{1}(-n,\sigma,x)=\sum_{p=0}^{n}\frac{(-1)^{p}n!}{(n-p)!p!}\frac{\Gamma(\sigma)}{\Gamma(\sigma+p)}x^{p}\,,
\end{eqnarray}
The relation between the Laguerre polynomials and confluent
hypergeometric functions as
$L_{n}^{\eta}(x)=\frac{\Gamma(n+\eta+1)}{n!\Gamma(\eta+1)}\,_{1}F_{1}(-n,\eta+1,x)$
[21] gives the wave functions as
\begin{eqnarray}
F(r)=\frac{n!\Gamma(2a_{1}+1)}{n+2a_{1}+1}\,r^{a_{1}+\frac{1}{2}}e^{-\,\epsilon_{n\kappa}
r}L_{}^{(2a_{1})}(2\epsilon_{n\kappa} r)\,.
\end{eqnarray}
Inserting the parameters in Eq. (7) into Eq. (12), we obtain the
energy spectra of the Kratzer potential in the presence of a
Coulomb-type tensor interaction
\begin{eqnarray}
E_{n\kappa}=m_{0}\frac{\left[1+2n+\sqrt{(1+2\kappa)^2+4C(1+2\kappa+C)+8Da^2(E_{n\kappa}+m_{0})\,}\right]^2-16D^2a^2}
{\left[1+2n+\sqrt{(1+2\kappa)^2+4C(1+2\kappa+C)+8Da^2(E_{n\kappa}+m_{0})\,}\right]^2+16D^2a^2}\,.
\end{eqnarray}
which is the same with the ones obtained in Ref. \cite{or} for
$C=0$. In Table 1, we present the numerical results for the case
where $C\neq0$ to see the effect of the tensor interaction on the
energy spectrum. It is seen that the energy eigenvalues decrease
while the strength of the tensor interaction increases. It is
interesting to obtain the solution for the Coulomb case, setting
the second term in Eq. (1) to zero, where we set the parameters as
$Da=Ze^2$ ($Ze^2$ is the charge of the nucleus) and $\kappa=\ell$
in Eq. (18). This gives in the absence of tensor potential
($\hbar=c=1$)
\begin{eqnarray}
E_{n\ell}=m_{0}\,\left(1-\frac{8Z^2e^4}{(n+\ell+1)^2+4Z^2e^4}\right)\,.
\end{eqnarray}
It is worth to say that this result is the same with the one
obtained for the well-known Dirac-Coulomb problem in Ref.
\cite{kth}.

Now we intend to give the results for the Kratzer potential and
Coulomb potential in the non-relativistic limit.
\subsection{Non-Relativistic Kratzer Limit} In order to obtain
the energy eigenvalues of the Kratzer potential in the
non-relativistic limit for $C=0$, we set
$E_{n\kappa}-m_{0}c^2\simeq E_{n\ell}$ and
$E_{n\kappa}+m_{0}c^2\simeq 2m_{0}c^2$. Using this assumptions and
with the help of Eq. (7),  Eq. (12) gives ($\hbar=c=1$)
\begin{eqnarray}
E_{n\ell}=-\frac{8D^2a^2m_{0}}{\left[1+2n+\frac{1}{2}\sqrt{1+4\ell(\ell+1)+16Dm_{0}a^2\,}\right]^2}\,.
\end{eqnarray}
which is exactly same with the result obtained in Ref. \cite{cb}.
\subsection{Non-Relativistic Coulomb Limit} We expand Eq. (19)
into a series in terms of the parameter $Ze^2$ in order to obtain
the energy spectrum of the Coulomb potential in the
non-relativistic limit in the absence of tensor interaction. So,
we find the energy equation for $Ze^2\ll 1$
\begin{eqnarray}
E_{n\ell} \sim
m_{0}\left[1-\frac{8Z^2e^4}{(n+\ell+1)^2}+\frac{32Z^4e^8}{(n+\ell+1)^4}+\ldots\right]\,.
\end{eqnarray}
which is in agreement with the results obtained in literature
\cite{kth}.

Finally, we briefly study the case where the scalar and vector
potentials are equal magnitude but different sign, \textit{i.e.},
$S(r)=-V(r)$ in which the Dirac equation has pseudo-spin symmetry
\cite{rlh,cyf,ad1}. For convenience for the rest of computation,
we give the differential equation satisfying by the lower
spinor-component. Following the same steps, we obtain the
following
\begin{eqnarray}
\frac{d^2\phi(r)}{dr^2}+\frac{1}{r}\frac{d\phi(r)}{r}-\frac{1}{r^2}\left(A^2_{1}+A^2_{2}r+
\varepsilon^2_{n\kappa}r^2\right)\phi(r)=0\,,
\end{eqnarray}
where
\begin{subequations}
\begin{align}
A^2_{1}&=\kappa(\kappa-1)+C(2\kappa+C-1)-2Da^2(m_{0}-E_{n\kappa})+\frac{1}{4}\,,\\
A^2_{2}&=4Da(m_{0}-E_{n\kappa})\,,\\
\varepsilon^2_{n\kappa}&=m^2_{0}-E^2_{n\kappa}=\epsilon^2_{n\kappa}\,.
\end{align}
\end{subequations}
The wave functions can be written as
\begin{eqnarray}
G(r)=\frac{n!\Gamma(2A_{1}+1)}{n+2A_{1}+1}\,r^{A_{1}+\frac{1}{2}}e^{-\,\varepsilon_{n\kappa}
r}L_{}^{(2A_{1})}(2\varepsilon_{n\kappa} r)\,,
\end{eqnarray}
and the energy eigenvalue equation of the Kratzer potential with a
tensor interaction is written as
\begin{eqnarray}
E_{n\kappa}=m_{0}
\left[1-\frac{2\left[1+2n+\sqrt{(1-2\kappa)^2+4C(2\kappa+C-1)+8Da^2(E_{n\kappa}-m_{0}c^2)\,}\right]^2}
{\left[1+2n+\sqrt{(1-2\kappa)^2+4C(2\kappa+C-1)+8Da^2(E_{n\kappa}-m_{0}c^2)\,}\right]^2+16D^2a^2}\right]\,.\nonumber\\
\end{eqnarray}
which is the same with the ones obtained in Ref. \cite{or} $C=0$.
Our numerical results are given for the pseudo-spin symmetric case
in Table 1 for $C\neq 0$ and see that the energy eigenvalues
increase while the strength of the tensor interaction increases.

\section{Conclusion}
We have exactly solved the radial Dirac equation for the Kratzer
potential in the presence of a tensor interaction term by using
the Laplace transform approach. The energy eigenvalues and the
corresponding upper- and lower-spinor components of the potential
are computed for the cases of the spin and pseudo-spin symmetry.
We have also obtained the energy eigenvalues of the Coulomb
potential by choosing suitable parameter values. We have also
studied the non-relativistic limit for the above potentials.

\section{Acknowledgments}
This research was partially supported by the Scientific and
Technical Research Council of Turkey. The authors would like to
thank the referee for suggestions improving the manuscript
greatly.

%\bibliographystyle{phaip}
%\bibliographystyle{unsrt}
%\bibliography{deneme}

\newpage

\begin{table}
\caption{Energy eigenvalues of the Kratzer potential in the
presence of the tensor interaction for the parameter values as
$m_{0}=5$ fm$^{-1}$, $D=1.25$ fm$^{-1}$, $a=0.35$ fm$^{-1}$
\cite{or}.}
\begin{center}
\begin{tabular}{cccc}
&  &spin symmetry  & \\ \hline $n$ & $\kappa$ & $C=0.25$ &
$C=0.50$ \\
0 & -2 & 4.63346 & 4.57251 \\
  & -3 & 4.79992 & 4.76888 \\
  & -4 & 4.88031 & 4.86515 \\
  & -5 & 4.92171 & 4.91357 \\
1 & -2 & 4.82214 & 4.80240 \\
  & -3 & 4.88546 & 4.87236 \\
  & -4 & 4.92329 & 4.91564 \\
  & -5 & 4.94578 & 4.94114 \\
\hline
&  &pseudo-spin symmetry  & \\
\hline $n$ & $\kappa$ & $C=0.25$ & $C=0.50$ \\
1 & -1 & -4.46315 &  \\
  & -2 & -4.83578 & -4.80233 \\
  & -3 & -4.90668 & -4.89445 \\
  & -4 & -4.93862 & -4.93245 \\
2 & -1 & -4.75526 &  \\
  & -2 & -4.90136 & -4.88615 \\
  & -3 & -4.93728 & -4.93064 \\
  & -4 & -4.95581 & -4.95208 \\
\end{tabular}
\end{center}
\end{table}

\end{document}